\documentclass[12pt]{article}
\usepackage{makeidx,amsfonts,graphicx,amsthm}
\newtheorem{theorem}{Theorem}{\bfseries}{\rmfamily}
\newtheorem{lemma}[theorem]{Lemma}{\bfseries}{\rmfamily}
\newtheorem{proposition}[theorem]{Proposition}{\bfseries}{\rmfamily}
{\bfseries}{\rmfamily}
\theoremstyle{definition}
\newtheorem*{definition}{{Definition}}{\bfseries}{\rmfamily}

\newtheorem{algorithm}[theorem]{Algorithm}{\bfseries}{\rmfamily}
\newcommand{\bt}{\begin{tabular}}
\newcommand{\et}{\end{tabular}}
\newcommand{\ba}{\begin{array}}
\newcommand{\ea}{\end{array}}
\newcommand{\bqa}{\begin{eqnarray*}}
\newcommand{\eqa}{\end{eqnarray*}}
\newcommand{\ds}{\displaystyle}

\def\rr{{\mathbb R}}
\def\zz{{\mathbb Z}}
\def\nn{{\mathbb N}}
\def\ff{{\mathbb F}}
\def\fqw{\left(\ff_q^M\right)^\infty}
\def\mod{{\mbox {\rm\ mod\ }}}
\def\hx{\hspace{-3 mm}}
\newcommand{\bthm}{\begin{theorem}}
\newcommand{\ethm}{\end{theorem}}
\newcommand{\bpro}{\begin{proposition}}
\newcommand{\epro}{\end{proposition}}
\newcommand{\bde}{\begin{definition}}
\newcommand{\ede}{\end{definition}}

\begin{document}

\pagestyle{plain}

\title{\bf
The Asymptotic Normalized Linear Complexity of Multisequences\\}

\author{Michael Vielhaber\footnote{
Supported by grant FONDECYT  1040975 of
CONICYT, Chile
}
\and M\'onica del Pilar Canales Chac\'on\footnotemark[\value{footnote}]
\\\\
Instituto de Matem\'aticas\\ 
Universidad Austral de Chile\\
Casilla 567\\
Valdivia\\\\
Phone/Fax\ \ \  ++56 / 63 / 221298
\\\\
{\rm\{}{\tt vielhaber, monicadelpilar{\rm\}}\ @\ gmail.com}
}

\date{}

\maketitle

\newpage

\begin{abstract}

We show that the asymptotic linear complexity of a multisequence
$a\in \fqw$ that is 
$I := \liminf_{n\to\infty}\frac{L_a(n)}{n}$ and 
$S := \limsup_{n\to\infty}\frac{L_a(n)}{n}$  satisfies the inequalities
$$\frac{M}{M+1}\leq S\leq 1\mbox{\rm \  and\ } M(1-S)\leq I \leq
1-\frac{S}{M},$$  
if all $M$ sequences have nonzero discrepancy infinitely often,  
and all pairs $(I,S)$ satisfying
these conditions are met by $2^{\aleph_0}$  multisequences~$a$.
This answers an Open Problem by Dai, Imamura, and Yang.

{\bf Keywords:} Linear complexity, multisequence, Battery Discharge
Model, isometry.

\end{abstract}

\newpage

\section{Introduction}

Given $M$ formal power series 
\[G_m = \sum_{t=1}^\infty
a_{m,t}x^{-t}\in \ff_q[[x^{-1}]], 1\leq m \leq M \mbox{\rm\ with\ } a =
(a_{m,t})\in \fqw,\]  
the linear complexity $L_a(n) $ is
defined as the smallest degree $\deg(v)$ of a denominator polynomial
$v\in\ff_q[x]$,
which approximates all $G_m$'s up to $x^{-n}$:
\[ \exists u_1,\dots,u_M\in\ff_q[x]\colon G_m=\frac{u_m(x)}{v(x)} +
o(x^{-n}).\] 

Typically $L_a(n) \approx n\cdot \frac{M}{M+1}$, and we define the
{\it linear   complexity  deviation}
\[d := d_a(n) := L_a(n) -\left\lceil\frac{M}{M+1}\cdot n\right\rceil.\]

In Section 2, we recall Dai and Feng's \cite{Dai} multi--Strict Continued
Fraction Algorithm (mSCFA) and our  Battery--Discharge--Model
(BDM) \cite{BDM}\cite{towards},  
which keeps track of the  linear complexity deviation of {\it
  all} multisequences in $\fqw$ simultaneously.

The normalized linear complexity is defined as 
$\overline L_a(n) =  {L_a(n)}/{n}$ with $0\leq \overline L_a(n) \leq
1$,  typically $\overline L_a(n) \approx M/(M+1)$,
similarly the normalized deviation
$\overline d_a(n) =  {d_a(n)}/{n}$ is typically $\overline d\approx 0$, and in
Section 3, we show bounds for the possible values for 
$I := \liminf\frac{L_a(n)}{n}$ and 
$S := \limsup\frac{L_a(n)}{n}$.

In Section 4 we give an algorithm to construct an
$M$--multisequence  (over
any finite field) with any allowed parameters $I,S$.

The final Section 5 considers the cardinality, Hausdorff
dimension, and measure of the set of multisequences matching a given pair
$(I,S)$. Nieder\-rei\-ter and Wang \cite{NW,NW1,WN} recently have shown
that with measure one we have $I=S=M/(M+1)$.
We shall see however  that
all the other points $(I,S)$ matching the conditions are also  met by 
$2^{\aleph_0}=\left|\left(\ff_q^M\right)^\infty\right|$ sequences $a$,
leading to a set of positive Hausdorff dimension at least for $S<1$.

This answers an Open Problem posed by Dai, Imamura, and Yang
\cite{DIY}, and extends the work in \cite{DIG} for $M=1$ to arbitrary
parallelism $M$.

\section{Diophantine Approximation of Multisequences}

We start with  the multi--Strict Continued Fraction Algorithm (mSCFA) by
Dai and Feng \cite{Dai}. The mSCFA calculates a best
simultaneous  approximation to a set of $M$ formal power series
$G_m = \sum_{t=1}^\infty a_{m,t}x^{-t}\in\ff_q[[x^{-1}]],$ 
$\ 1\leq m\leq M.$
It computes  a sequence $(u_m^{(m,n)}/v^{(m,n)})$
of  approximations in $\ff_q(x)$  in the order
$(m,n) = (M,0),(1,1),(2,1),\dots,(M,1)(1,2,),(2,2),\dots$ with 
$$G_k = \sum_{t\in\nn}a_{k,t}\cdot x^{-t} =
\frac{u_k^{(m,n)}(x)}{v^{(m,n)}(x)}+o(x^{-n}),
\forall\ 1\leq m,k\leq M, \forall n\in\nn_0.$$ 
We will denote the degree of $v^{(m,n)}(x)$ by $\deg(m,n)\in\nn_0$ instead of
$d$ as in~\cite{Dai} (we will use $d$ differently).
Then the multisequence $a=(a_{m,n})\in\fqw$ has the linear complexity profile 
 $(\deg(M,n))_{n\in\nn_0}= (L_a(n))_{n\in\nn_0}.$

The mSCFA also uses $M$ auxiliary degrees $w_1,\dots,w_M\in\nn_0$.
The update of these values depends on a so--called ``{\it discrepancy}'' 
$\delta(m,n)\in\ff_q$.
$\delta(m,n)$ is zero if the current approximation predicts 
correctly the value $a_{m,n}$, and $\delta(m,n)$ is nonzero otherwise.
Furthermore, the polynomials $u_m(x)$ and $v(x)$ 
are updated, crucial
for the mSCFA, but of no importance for our concern.

\begin{algorithm}
{\tt mSCFA}\ \cite{Dai}\\
$\deg:=0; w_m:=0,1\leq m\leq M$\\
{\tt FOR} $n := 1,2,\dots$\\
\hspace*{6 mm}    {\tt FOR} $m:=1,\dots,M$\\
\hspace*{12 mm}        {\tt compute} $\delta(m,n)$\ \ //discrepancy\\
\hspace*{12 mm}        {\tt IF} $\delta(m,n) = 0$: \{\} // do nothing, 
\cite[Thm.~2, Case 2a]{Dai}\\
\hspace*{12 mm}        {\tt IF} $\delta(m,n) \neq 0$ {\tt AND} $n-\deg-w_m\leq 0: \{\}$
 // \cite[Thm.~2, Case 2c]{Dai}\\
\hspace*{12 mm}        {\tt IF} $\delta(m,n) \neq 0$ {\tt AND} $n-\deg-w_m > 0$: 
 // \cite[Thm.~2, Case 2b]{Dai}\\
\hspace*{18 mm}          $\mbox{deg\_copy}:=\deg$\\
\hspace*{18 mm}          $\deg := n -w_m$\\
\hspace*{18 mm}          $w_m := n - \mbox{deg\_copy}$\\
\hspace*{12 mm}     {\tt ENDIF}\\
\hspace*{6 mm}     {\tt ENDFOR}\\
{\tt ENDFOR}
\end{algorithm}

The linear complexity grows like $\deg(M,n)\approx \left\lceil
n\cdot \frac{M}{M+1}\right\rceil$  (exactly, if always
$\delta(m,n)\neq 0$), and 
the $w_m\approx\left\lfloor\frac{n}{M+1}\right\rfloor$.
We therefore extract the {\it deviation} from this average behaviour as
\begin{eqnarray}
d &:=& \deg - \left\lceil n\cdot\frac{M}{M+1}\right\rceil,
\end{eqnarray}
the linear complexity deviation or degree deviation, 
which we call the ``{\it drain}'' value,  and
\begin{eqnarray}
 b_m &:=&  \left\lfloor n\cdot\frac{1}{M+1}
  \right\rfloor-w_m,\ \ \ 1\leq m\leq M,
\end{eqnarray}
the deviation of the auxiliary degrees, which we call the  ``{\it
  battery charges}''. 

We establish the behaviour of $d$ and $b_m$ in two steps. First we
treat the change of $d,b_m$ when increasing $n$ to $n+1$ 
(keeping $\deg,w_m$ fixed for the moment):
\begin{equation}
\deg - \left\lceil (n+1)\cdot\frac{M}{M+1}\right\rceil=\left\{
\ba{ll}
 \deg - \left\lceil n\cdot\frac{M}{M+1}\right\rceil-1,&n \not\equiv M
 \mod (M+1),\\ 
 \deg - \left\lceil n\cdot\frac{M}{M+1}\right\rceil,  &n \equiv M \mod (M+1),\\
\ea 
\right.
\end{equation}
and
\begin{equation}
\left\lfloor (n+1)\cdot\frac{1}{M+1}\right\rfloor-w_m=\left\{
\ba{ll}
\left\lfloor n\cdot\frac{1}{M+1}  \right\rfloor-w_m,  &n \not\equiv M
\mod (M+1),\\ 
\left\lfloor n\cdot\frac{1}{M+1}  \right\rfloor-w_m+1,&n \equiv M \mod (M+1).\\
\ea
\right.
\end{equation}

Hence, by $(3)$ we have to decrease $d$ in all steps, except when 
$n\equiv M \to$\linebreak
$n\equiv 0 \mod (M+1)$, 
and only here we increase all $M$ battery values $b_m$, by~$(4)$.

With $d(M,0)=b_m(M,0):=0,\forall m$,  initially, we obtain the invariant
\begin{eqnarray}
&&d(M,n) + \left(\sum_{m=1}^M b_m(M,n)\right)+ n\mod (M+1)=0.
\end{eqnarray}
Now, for $n$ fixed, the $M$ steps of the inner loop of the mSCFA
change $w_m$ and $\deg$  only in the case of $\delta(m,n)\neq 0$ and 
$n-\deg-w_m>0$ that is 
\[n-\deg-w_m>0\Leftrightarrow 
n-\left(d+ \left\lceil n\cdot\frac{M}{M+1}\right\rceil\right) 
-\left( \left\lfloor n\cdot\frac{1}{M+1} \right\rfloor-b_m\right) >0\] 
$\Leftrightarrow b_m >d.$
In this case $\delta\neq 0$ and $b_m>d$, the new values are
(see mSCFA)
\begin{eqnarray}
\deg^+=n-w_m&\mbox{\rm \ and \ }&w_m^+ = n-\deg
\end{eqnarray}
and thus in terms of the BDM variables:
\bqa
d^+ \stackrel{(1;6)}{=} 
(n-w_m) - \left\lceil\frac{n\cdot M}{M+1}\right\rceil
\stackrel{(2)}{=} 
\left\lfloor\frac{n}{M+1}\right\rfloor + b_m 
- \left\lfloor\frac{n}{M+1}\right\rfloor = b_m
\eqa
and
\bqa
b^+_m \stackrel{(2;6)}{=} 
\left\lfloor\frac{n}{M+1}\right\rfloor - n + \deg
\stackrel{(1)}{=} 
-\left\lceil\frac{n\cdot M}{M+1}\right\rceil+ \left(d 
+\left\lceil\frac{n\cdot M}{M+1}\right\rceil\right)
=d,
\eqa
an interchange of the values $d$ and $b_m$. We say in this case that
``battery $b_m$ {\it discharges} the excess charge into the drain''. 
A discharge does not affect the invariant $(5)$, which is thus valid
for every timestep $(m,n)$.

In the limit, as  $n\to\infty$, we  want to obtain $d$  as a probability 
distribution over  {\it all} multisequences in $\fqw$. Since we do not actually compute 
with a given multisequence $a\in\fqw$, we have to model the distinction between
$\delta=0 $ and $\delta\neq 0$ probabilistically.

\begin{proposition}
In any given position $(m,n), 1\leq m\leq M, n\in\nn$ of the formal
power series, 
exactly one choice for the next symbol  $a_{m,n}$
will yield a discrepancy $\delta(m,n)=0$, all other $q-1$ symbols from $\ff_q$ 
result in some  $\delta(m,n)\neq 0$.
\end{proposition}

\begin{proof}\ 
The current approximation $u_m^{(m,n)}(x)/v^{(m,n)}(x)$ determines
exactly  {\it one} approximating coefficient sequence  for the $m$--th
formal power series $G_m$. 
The (only) corresponding symbol leads to $\delta=0$, all other symbols
lead to $\delta\neq 0$.
\end{proof}

In fact, for every position 
$(m,n)$, each discrepancy value $\delta(m,n) \in\ff_q$
occurs exactly once for some  $a_{m,n}\in\ff_q$, 
in other words (see \cite{CV,VSETA} for $M=1$):
\\\\
{\bf Fact}
\ \ 
{\it The  mSCFA induces an isometry on $\fqw$.
}
\\\\
Hence, we can model $\delta=0$ as occurring with
probability $1/q$, and  $\delta\neq 0$ as having probability $(q-1)/q$.
In terms of $d,b_m$, we have the following equivalent
probabilistic formulation of the mSCFA:

\newpage
\begin{algorithm}
{\tt Battery-Discharge-Model BDM} (probabilistic mSCFA)\\
$d:=0; b_m:=0,1\leq m\leq M$\\
{\tt FOR} $n := 1,2,\dots$\\
\hspace*{6 mm}  {\tt IF} $n\equiv M \mod M+1:\\
\hspace*{12 mm}      b_m := b_m + 1, 1\leq m \leq M$\\
\hspace*{6 mm}   {\tt ELSE}\\ 
\hspace*{12 mm}       $d := d -1$\\ 
\hspace*{6 mm}     {\tt ENDIF}\\
\hspace*{6 mm}    {\tt FOR} $m:=1,\dots,M$\\
\hspace*{12 mm}        {\tt IF} $b_m > d$:\\
\hspace*{18 mm}            {\tt WITH} prob.~$(q-1)/q$:\\
\hspace*{24 mm}                {\tt swap}$(b_m,d)$  // Discharge of
battery $b_m$\\
\hspace*{18 mm}            {\tt WITH} prob.~$1/q$:\\
\hspace*{24 mm}                \{\}  // Do nothing, since $\delta=0$\\
\hspace*{12 mm}        {\tt ELSE}\\
\hspace*{18 mm}            \{\}  // Do nothing, since $b_m\leq d$\\
\hspace*{12 mm}         {\tt ENDIF}\\
\hspace*{6 mm}     {\tt ENDFOR}\\
{\tt ENDFOR}\\\\
\end{algorithm}

\section{Normalized Linear Complexity I:\\ 
Bounds for {$\liminf$} and {$\limsup$}}

We need the following facts about the mSCFA and BDM:

\noindent 1. $\ds 0\leq {L_a(n)}/{n}\leq 1$.\\
2. The invariant $(5)$.\\
3. Proposition 2 that is, after each prefix we can enforce
both $\delta(m,n)=0$ and $\delta(m,n)\neq 0$ by choosing an
appropriate $a_{m,n}$, for any finite field $\ff_q$.

\bde {\it Asymptotic Normalized Bounds}\\
We denote the asymptotic lower bound for the normalized linear
complexity  by 
\bqa
I&:=& \liminf_{n\to \infty} \frac{L_a(n)}{n}
= \liminf_{n\to \infty} \frac{\deg(M,n)}{n}
\eqa
and for the normalized drain or linear complexity deviation by
\bqa
\tilde I&:=& \liminf_{n\to \infty} \frac{d(M,n)}{n}
= I- \frac{M}{M+1},
\eqa
similarly the asymptotic upper bounds are
$$
S:= \limsup_{n\to \infty} \frac{L_a(n)}{n}
\mbox{\rm \ \ and\ \ }
\tilde S:= \limsup_{n\to \infty} \frac{d(M,n)}{n}
= S- \frac{M}{M+1}.
$$
\ede

\bde {\it Active Series}\\
We call a formal power series $G_m$  {\it active}, if  
$\delta({m,n})\neq 0$ infinitely
often and denote the number of active series by 
$K$\ $(0\leq K\leq M)$.
\ede

\bpro
$K$ is the number of $\ff_q(x)$--independent 
irrational series that is 
$$K = dim_{\ff_q(x)}<1,G_1,\dots,G_M> - 1.$$
\epro

{\it Proof.}
If the discrepancy sequence of  a series is ultimately zero, this
series will be 
either rational or dependent (as $\ff_q(x)$--linear combination) on the
active series. Thus $K$ is the number of $\ff_q(x)$--independent 
irrational series, where including 1 as generating element of the
vector space, and decrementing the dimension
removes any effect of ultimately periodic (rational) series.$\Box$

Since nonactive series do not change the linear complexity profile,
we shall in fact assume for the purpose of deriving bounds
that all $M$ series are active.
After proving a technical lemma, we will obtain bounds for $I$, $S$,
$\tilde I$, and $\tilde S$, which will turn out to be tight in the
next section.

\begin{lemma}
$(i)$ If $G_m$ is active, and if there is an $n_0$ with $\tilde I \leq
  {d(m,n)}/{n}\leq 
  \tilde S$ for all $n \geq n_0$, then there is also 
an $n_1$ with $ \tilde I \leq  {b_m(m,n)}/{n}\leq  \tilde S$ for all
  $n\geq n_1$. 

$(ii)$ Asymptotically, the normalized drain and batteries sum up to
zero,\linebreak
$\lim_{n\to\infty} d(m,n)/n+\sum_{k=1}^M (b_k(m,n)/n) = 0,\forall
1\leq m\leq M.$
\end{lemma}

{\it Proof.}
$(i)$ 
Let $n_1$ be the first time after $n_0$ where $b_m$ discharges
(since $G_m$ is active, such an $n_1$ exists).
Then, we have 
$\tilde I\leq d(m,n_1)/n_1  \leq \tilde S$ by assumption, and also 
$\tilde I\leq d^+(m,n_1)/n_1 =  b_m(m,n_1)/n_1 \leq \tilde S$ after
the discharge.  The same holds for every $n^*>n_1$ where $b_m$
discharges and as $G_m$ is active, infinitely many such $n^*$ exist.  
Also, between $n_1$ and $n^*$, $b_m/n$ has to stay between $\tilde I $
and $\tilde S$ since otherwise it would make $d/n$ leave this interval
at discharge. Hence, not only $d/n$, but all $b_m/n$ for  active batteries
$b_m$ are  eventually bounded by $\tilde I$ and $\tilde S$.

$(ii)$ Since $(n \mod(M+1))/n \to 0$, this follows from the invariant
$(5)$.$\Box$

\bthm
Let $a\in \fqw$ with $\delta({m,n})\neq 0$ infinitely often for all
$1\leq m\leq M$ $($all series active$)$.
Then $I,S,\tilde I,\tilde S$  satisfy conditions
\begin{equation}
\ba{cccccccccccccc}
&  \ds\frac{M}{M+1} &\leq& S &\leq& 1,          &0&\leq& \tilde S&\leq&
\ds\frac{1}{M+1},\\
{\mbox \rm  and}\\
&  M(1-S)     &\leq& I &\leq& 1-\ds\frac{S}{M},\ \ \ \ &-M\cdot \tilde S&\leq&
\tilde I &\leq&\ds -\frac{\tilde S}{M}.
\ea
\end{equation}
\ethm

{\it Proof.}  We show the four inequalities in turn:

\noindent
a) $S\leq 1$ or $\tilde S \leq 1/(M+1)$:

Since $L_a(n) \leq n$, the normalized linear complexity stays below or
at 1, and the normalized drain below or at $\tilde S \leq 1-\frac{M}{M+1} =
\frac{1}{M+1}$. 
\\\\
b) $M/(M+1)\leq S$ or $0\leq \tilde S$:

The maximum of the $b_m$ and $d$  is larger than or equal to 
the average over all $b_m$
and $d$, which is zero. From time to time, $d$ assumes this maximum
after discharging the currently largest $b_m$ (all $G_m$ are active).
Hence $\tilde S \geq 0$ and $S \geq \frac{M}{M+1}$.
\\\\
c) $M(1-S)\leq I$ or $-M\cdot \tilde S\leq \tilde I$:

For all $\varepsilon > 0$ and $n\geq n_1$ for some $n_1$,   
$b_m/n \leq \tilde S+\varepsilon$. 
So $\sum_m b_m/n \leq M\cdot(\tilde S+\varepsilon)$, 
and with $d/n+\sum_m b_m/n \to 0$ (Lemma 7 (ii)), 
we have $d/n \geq -M\cdot (\tilde S+\varepsilon)$.
With $n\to\infty, \varepsilon\to 0$, therefore  $\tilde I\geq -M\cdot
\tilde S$.  
Now, $\ds I = \tilde I +\frac{M}{M+1} \geq -M\cdot
\left(S-\frac{M}{M+1}\right)+\frac{M}{M+1} = (M+1)\frac{M}{M+1}-M\cdot
S=M(1-S) $.  
\\\\
d) $I\leq 1-S/M$ or $\tilde I \leq -\tilde S/M$

Asymptotically, the drain and all (active) batteries stay above $I$ by
Lemma $5(i)$. The normalized values thus satisfy
\[\forall \varepsilon_1>0,\exists n_1,\forall n>n_1,
\forall m,\forall k\colon\ 
d(m,n)/n\geq \tilde I-\varepsilon_1, b_k(m,n)/n\geq \tilde
I-\varepsilon_1.\]
Also, there are infinitely many timesteps where the normalized drain
value $d/n$ is arbitrarily near $\tilde S$ after a discharge. Some
battery, $b_{m^*}$ say, is involved in infinitely many of these
discharges and hence itself was near $\tilde S$ before those discharges:
\[\forall \varepsilon_2>0,\exists m^*,
\forall n, \exists n_1>n\colon\  
{b_{m^*}(m^*,n_1)}/{n_1}>\tilde S-\varepsilon_2.\]

\noindent  
Therefore, at the infinitely many timesteps $(m^*,n_1)$, we have  with
Lemma~$5(ii)$: 
$$0\leftarrow \frac{b_{m^*}(m^*,n_1)}{n_1} + \frac{d(m^*,n_1)}{n_1} +
\sum^M_{\scriptsize
\ba{l}
k=1\\ 
k\neq m^*\hx
\ea
}
\frac{b_{k}(m^*,n_1)}{n_1} \geq (\tilde S-\varepsilon_2)+(1+(M-1))(\tilde
I-\varepsilon_1).$$ 
Letting $n\to\infty$ and $\varepsilon_1,\varepsilon_2\to 0$ gives 
$0\geq \tilde S+M\cdot \tilde I\Longleftrightarrow
\tilde I \leq -\frac{\tilde S}{M},$
and thus 
$$I = \tilde I +\frac{M}{M+1} 
\leq -\frac{\tilde S}{M} +\frac{M}{M+1}  
= -\frac{S-\frac{M}{M+1}}{M} +\frac{M}{M+1}
=1 -\frac{S}{M}.$$  

\vspace*{- 5 mm}\hfill$\Box$

Now, again incorporating the possibility of inactive sequences, we may
state as a corollary:
\bthm
For any multisequence $a\in \fqw$,
the bounds
$I$,
$S$,
$\tilde I$,
$\tilde S$ satisfy
\begin{equation}
\ba{cccccccccccccc}
&  \ds\frac{K}{K+1} &\leq& S &\leq& 1,          &0&\leq& \tilde S&\leq&
\ds\frac{1}{K+1}\\
{\mbox \rm  and}\\
&  K(1-S)     &\leq& I &\leq& 1-\ds\frac{S}{K},\ \ \ \ &-K\cdot \tilde S&\leq&
\tilde I &\leq&\ds \frac{-\tilde S}{K}\\
\ea
\end{equation}
\noindent
for some $1\leq K\leq M$,\\ 
or $a$ is ultimately periodic, hence $K=0$, $I=S=0$, and $\ds\tilde I =
\tilde S = -\frac{M}{M+1}$.
\ethm

{\it Proof.}
If all series have ultimately periodic coefficient sequences, 
$L_a(n) = O(1)$ and thus $L_a(n)/n\to 0$.
Otherwise, apply Theorem 6 with $M:=K$, since the
$M-K$ inactive series asymptotically do not affect $L_a,\deg,$ or $d$. 
 \hfill$\Box$

We visualize all allowed pairs $(I,S)$ in Figure~1.
\nopagebreak

The allowed parameters lie on the point $(0,0)$ for $K=0$, on the line
$I+S=1,I\leq S$ for $K=1$, and on overlapping triangles with endpoints
$(0,1), (\frac{K}{K+1},\frac{K}{K+1})$ and $(\frac{K-1}{K},1)$ for
$2\leq K\in\nn$ ($K=0,\dots,5$ shown). The allowed area thus is not
convex, not even connected.
The points on the diagonal $I=S$ are just the values 
$(\frac{K}{K+1},\frac{K}{K+1}), K\in\nn_0$, for convergent normalized
complexities, and almost all multisequences (in the sense of Haar measure) 
can be found here \cite{NW,NW1,WN}.

For $M$ sequences in parallel, all cases $0\leq K \leq M$
are allowed (see (8)).

\includegraphics{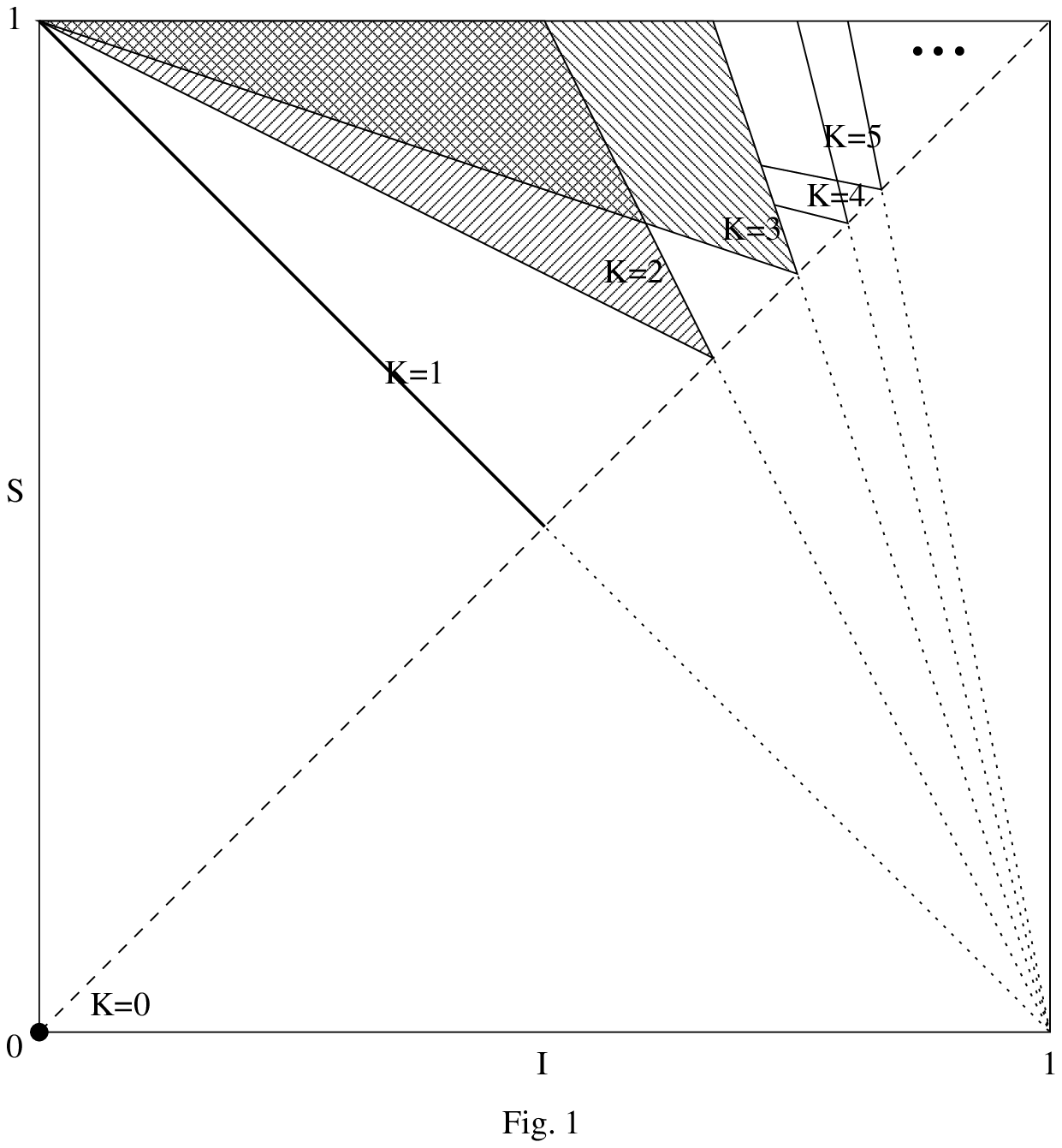}

\section{Normalized Linear Complexity II:\\ 
Existence of Multisequences\\ 
Meeting any Allowed {$\liminf$} and {$\limsup$}}

We next show that all pairs $(I,S)$ satisfying the conditions $(7)$,
resp.~$ (8)$
actually occur for some multisequence $a\in\fqw$, for any
finite field $\ff_q$. We construct a discrepancy sequence
$\delta(m,n)$ which leads to the specified behaviour of the normalized
linear complexity. From the sequence $\delta(m,n)$ one can then
obtain the actual coefficient sequence $(a_{m,n})$  applying the mSCFA.
We first assume  $K=M$ that is all sequences are active.

Since only the asymptotic behaviour is of importance, small effects 
from the integrality of all numbers can be ignored, and we assume from now
on $d,b_m\in\rr$. Also, we shall use $b_m(t), d(t)$ to mean $b_m(M,t),
d(M,t)$, since the precise internal timestep does not matter any longer.
The trajectories of the values for $d(t)$ and $b_m(t)$ shall follow a 
hexagon or butterfly pattern (see Figure 2).

\includegraphics{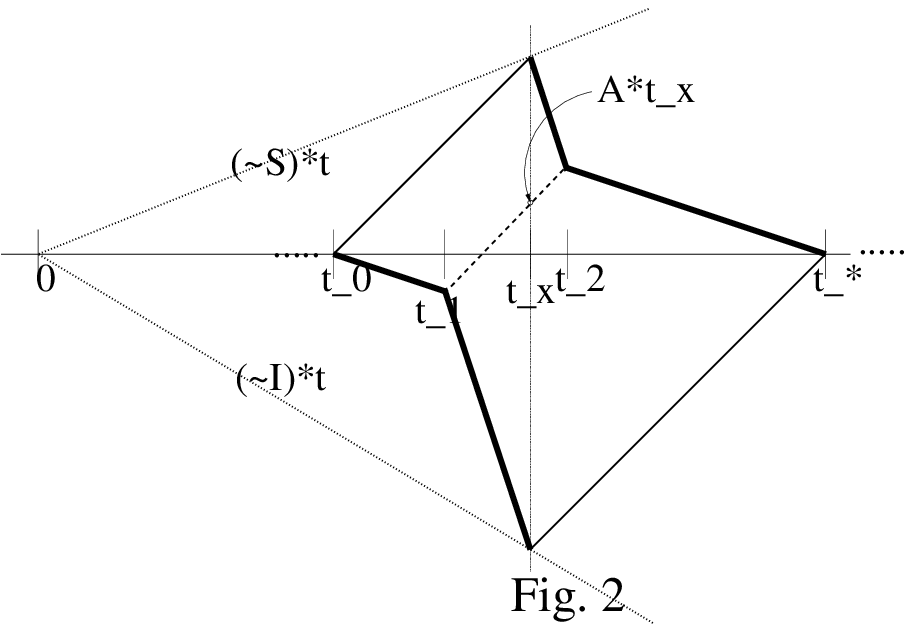}

\noindent
Drain $d$: { boldface},\\ 
Battery $b_1$: solid,\\ 
Batteries $b_2\dots b_M$: dashed (or ``buried'' in the $d$ trajectory),\\ 
Asymptotics $\tilde S\cdot t, \tilde I\cdot t$: dotted

The example shown uses the following values (for $A$ see (\ref{refA})) :

$M=K=3, t_0=96, S=0.85,\tilde S = 0.1, I=0.6,\tilde I = -0.15, A = 0,025$

$\ba{l|c|r|rcr|r|c}
t&t_0=96&t_1=132&t_x&=&160&t_2=172&t_*=256\\
\hline&&&&&&\\
d&0&-3&-24&\to&16&7&0\\
b_1&0&9&16&\to&-24&-15&0\\
b_2,b_3&0&-3&&4&&7&0\\\\
\ea$

{\it Description of the Construction}:\ 

We will stack an infinite sequence of these hexagonal patterns one
after the other, where each hexagon $H$ starts at time $t_0^{(H)}$ and
finishes at $t_*^{(H)}$. 

We consider 5 moments and 4 time intervals:

\noindent  At $t_0$, all batteries and the drain are at zero (this is always
possible for $t_0 := M$, with all discrepancies nonzero up to this
point). 

\noindent  $(t_0,t_1)$: $b_1$ grows ($\delta=0$), while $b_2=\cdots=b_M=d$ by
discharging ($\delta\neq 0$).

\noindent  At $t_1$, batteries $b_2,\dots,b_M$ stop to discharge.

\noindent  $(t_1,t_x)$: All batteries grow ($\delta\neq 0$).

\noindent  At $t_x$, battery $b_1$ has reached the value $\tilde S\cdot t_x$, 
while $d$ is at value $\tilde I\cdot t_x$. Now $b_1$ discharges, and 
thus $d$ becomes $\tilde I\cdot t_x$. It is at these points $t_x$, 
where $d$ assumes both limiting values and thus assures the asymptotic 
behaviour.

\noindent  $(t_x,t_2)$: All batteries are less than $d$ and thus inhibited to
discharge, irrespective of $\delta$.

\noindent  At $t_2$,  $d=b_2=\cdots=b_M$.

\noindent  $(t_2,t_*)$: All batteries except $b_1$ have to discharge, $\delta\neq 0$,
to ensure $b_2=\cdots=b_M=d$.

\noindent  At $t_*$, again  all batteries and the drain are at zero.

How are the different timesteps related: 

$t_x$: Since  battery $b_1$ grows 
(all $\delta(1,n)=0$) with slope $1/(M+1)$ (by (4)) until
touching the asymptotical line  $\tilde S\cdot t$ in $t=t_x$, we have
$$(t_x-t_0)\cdot \frac{1}{M+1} = \tilde S\cdot t_x\Longleftrightarrow t_x
=\frac{t_0}{1-\tilde S(M+1)}.$$

$A$: We require 
$d(t_x) = \tilde I\cdot t_x$ and
$b_1(t_x) = \tilde S\cdot t_x$. 
Assuming $b_2=\dots=b_M$, we then have 
$\tilde I + \tilde S + (M-1)\cdot {b_2}/{t_x}=0$ from $(5)$,
and thus 
\begin{equation}
A\cdot t_x := b_m(t_x) = \frac{-\tilde I-\tilde S}{M-1}\cdot t_x,\ \
2\leq m\leq M.
\label{refA}
\end{equation}

$t_1$:
We reach the point $(t_x,A\cdot t_x)$ from $(t_0,0)$ following
batteries $b_2\dots b_m$:
\[
\ba{crcl}
&A\cdot t_x &=& -\frac{1}{M(M+1)}(t_1-t_0) + \frac{1}{M+1}(t_x-t_1)\\
\Longleftrightarrow&t_x\cdot AM(M+1) &=& -t_1+t_x(1-\tilde S(M+1))+Mt_x-Mt_1\\
\Longleftrightarrow&t_1\cdot (M+1) &=& t_x(M+1)(1-\tilde S-AM)\\
\Longleftrightarrow&\ds  {t_1} &=& t_x\left(1- \tilde S-AM\right)
= \ds t_x\left(1+ \tilde I-A\right). 
\ea 
\]

$t_2$: 
Between $t_2$ and $t_x$, the initial difference $(\tilde S-a)t_x$
between $b_1$ and $b_2$ is overcome by $b_1$ with slope
$-\frac{M}{M+1}$ and $b_2$ with slope $\frac{1}{M+1}$, thus
$$(\tilde S-A)t_x=(t_2-t_x)\left(\frac{M}{M+1} + \frac{1}{M+1}\right)
\Leftrightarrow t_2=t_x(1+\tilde S-A).$$

$t_*$: The final time $t_*$ follows from
$$(t_*-t_0)\frac{1}{M+1} = (\tilde S-\tilde I)t_x
\Longleftrightarrow
t_*=t_0 + (M+1) \frac{(\tilde S - \tilde I)t_0}{1-\tilde S(M+1)}$$
by following the trajectory of $b_1$ with (always) slope $1/(M+1)$ by  
$(3)$.

The quotient $t_*/t_0$ is (excluding the case $S=1$, see Theorem~9 below)
$$\frac{t_*}{t_0} = \frac{1-\tilde I(M+1)}{1-\tilde S(M+1)}
=\frac{1-I}{1-S},$$
and we obtain a geometric progression
$$t_*^{(H-1)} = t_0^{(H)} = c_0\cdot\left(
\frac{1-I}{1-S}\right)^H$$
when  stacking hexagon $H$ directly  after hexagon $H-1$, $H\in \nn_0$,
starting in~$c_0$.

{\it Case $K<M$}:
Let now $0\leq K\leq M$. We construct a discrepancy sequence 
$\delta(m,n), n\in\nn, 1\leq m\leq K$, as before, which can be mapped via the
mSCFA to $K$ formal power series $G_1,\dots,G_K$ matching the bounds
$I$ and $S$.
The $M-K$ other formal power series are set  to $G_m=0, K+1\leq m\leq
M$,
not affecting the behaviour of $L_a$ or $d$.

\newpage
\begin{algorithm}

{\tt hexagon}

\noindent
{\tt INPUT} $\tilde I, \tilde S\in\rr$, $M\in\nn$\\
{\tt IF} $M=1$ {\tt THEN} $A:=\tilde I$ {\tt ELSE} $A := (-\tilde S - \tilde I)/(M-1)$\\
$t_1 := M+1$\\ 
$t_x := M+1$\\ 
$t=0$\\
{\tt FOREVER}\\
\hspace*{6 mm}    {\tt WHILE} ($t < t_1$)\\
\hspace*{12 mm}    $t$++\\
\hspace*{12 mm}    {\tt Do Not Discharge} $b_1$: $\delta(1,t) := 0$\\
\hspace*{12 mm}    {\tt Discharge} $b_2\dots b_M$: 
$\delta(m,t) :=1,\forall\ 2\leq m\leq M$\\
\hspace*{6 mm}    {\tt END}\\
\hspace*{6 mm}    {\tt WHILE} ($t < t_x$)\\
\hspace*{12 mm}    $t$++\\  
\hspace*{12 mm}    {\tt Do Not Discharge}: $\delta(m,t) := 0,
\forall\ 1\leq m\leq M$\\ 
\hspace*{6 mm}    {\tt END}\\
\hspace*{6 mm}    {\tt WHILE} ($\exists b_m\neq 0$)\ //
$t_x\dots t_2\dots t_*$\\
\hspace*{12 mm}    $t$++\\
\hspace*{12 mm}    {\tt Discharge All}: $\delta(m,t) :=
1,\forall\ 1\leq m\leq M$\\
\hspace*{6 mm}    {\tt END}\\
\hspace*{6 mm} //Optionally:\ Discharge All for $M+1$ additional timesteps \\
\hspace*{6 mm} //to obtain different multisequences  for the same $(I,S)$\\
\hspace*{6 mm}    $t_0 := t$\\
\hspace*{6 mm} $t_x := t_0/(1-\tilde S(M+1))$\ // $\tilde S \neq 1/(M+1),0$\\
\hspace*{6 mm} $t_1 := t_x(1+\tilde I-A)$\\ 
{\tt END}
\end{algorithm}

\bthm
Algorithm {\tt hexagon}  produces the discrepancy sequence  of a multisequence
$a\in\fqw$ with $\liminf L_a(n)/n=I$ and  
$\limsup L_a(n)/n=S$, provided that $I,S$ satisfy $(7)$. 
\ethm

{\it Proof.}
We already have shown by construction that the discrepancy sequence
produced by {\tt hexagon} corresponds to a multisequence $a\in\fqw$ with
asymptotic normalized linear complexities $I$ and $S$, provided $t\to\infty$.

It remains to be  verified that the algorithm indeed proceeds with
$t\to\infty$. This is {\it not} the case, only if $\tilde S=0$, hence 
$t_x =  t_0$, or for $\tilde S = 1/(M+1)$, leading to
$t_x=\infty$.
In these cases, {\tt hexagon} has to be adapted as follows: 
Instead of $\tilde S$, use $\hat S = 1/t_0$ or $\hat S = 1/(M+1) -1/t_0$, 
respectively, and otherwise
follow the same algorithm. Since $\hat S \to \tilde S$, 
we obtain the same asymptotics. \hfill $\Box$

\section{Cardinalities, Hausdorff Dimensions,\\ 
Measures}

Let ${\cal A}(I,S)\subset\fqw$ be the set of multisequences $a$ with 
asymptotic behaviour
$I = \liminf_{n\to\infty}{L_a(n)}/{n}$ and 
$S = \limsup_{n\to\infty}{L_a(n)}/{n}$.

{\it Cardinality}:\
For every admissible pair $(I,S)$, 
$|{\cal A}(I,S)|=2^{\aleph_0}=\left|\left(\ff_q^M\right)^\infty\right|$:
Between every $t_*$ and the next $t_0$, 
we may choose to include $M+1$ steps with $\delta(m,n)\neq 0$
(outcommented lines in Algorithm 8), 
leaving us again in $b_m=d=0,\forall m$.
Following immediately with the next hexagon would imply $\delta(1,n)=0$ at $b_1<0$,
leading to different multisequences.

{\it Measure}:\
Niederreiter and Wang \cite{NW,NW1,WN} recently have shown for all
$M\in\nn$ that 
$\mu\left({\cal A}(I,S)\right)=\left\{
\ba{cl}
1,&I=S= M/(M+1),\tilde I=\tilde S=0,\\
0,&\mbox{\rm otherwise}.\\
\ea\right.$

{\it Hausdorff dimension}:\
We map ${\cal A}(I,S)$  to the real unit
interval $[0,1]$ by 
$\iota:\fqw\ni a \mapsto \sum_{t=1}^\infty \sum_{m=1}^M
a_{t,m} \cdot q^{-(M\cdot(t-1)+m)}\in[0,1]\subset \rr$, where we
identify the set
$\ff_q$ with $\{0,1,\dots,q-1\}\subset \zz$ by some fixed
bijection, and denote its Hausdorff dimension by 
$D_H({\cal A}(I,S)) := D_H(\iota({\cal A}(I,S)))$.

\begin{theorem} $($Hausdorff Dimension$)$ 

Given a multiness $M$ and a pair $(I,S)$ of asymptotic limits, let
$K'$ be the largest $K\leq M$, such that $(I,S)$ lies within the
$K'$--th triangle 
$((0,1),$ $(\frac{K-1}{K},1),$ $(\frac{K}{K+1},\frac{K}{K+1}))$ 
$($or on the point $(0,0)$, or on the segment
$(0,1),(\frac{1}{2},\frac{1}{2})$ for $K'=0,1$, resp.$)$.

If no such $K'$ exists, $(I,S)$ is not admissible for  that $M$ and
${\cal A}(I,S)$ is empty. Otherwise the Hausdorff dimension of ${\cal
A}(I,S)$ within $\left(\ff_q^M\right)^\infty$ is  bounded by
\[\frac{K'}{M}\cdot\frac{1-S}{(M+1)(1-I)^2}\leq {\cal A}(I,S)\leq
\frac{K'}{M}.\] 
In particular, for $S < 1$ the Hausdorff dimension is positive.
\end{theorem} 

\begin{proof}
There may be at most $K'$ active sequences, since this is the largest
value permitted for $(I,S)$. We shall initially assume $M=K'$ and
later generalize to $M\geq K'$.

We define a subset of ${\cal A}(I,S)$ with discrepancy sequences that
alternate between hexagons according to Algorithm~8, $H_n,n\in \nn$
and ``fill'', $F_n,n\in\nn$, where the sequence may behave arbitrarily
while staying within the $(I,S)$ interval.

Assume that we have at least $(q^M)^{t_{N-1}\cdot (1-\frac{1}{N})}$
sequence prefixes up to $t_{N-1}$ (always possible for $N=1$
at $t_0=0$ with the single (empty) sequence $\varepsilon$).
We now want to append a hexagon. Since at the end of the fill phase, 
$d$ and the $b_m$ may be anywhere within $(I,S)$, we first discharge 
until $d=b_m=0$. This takes at most a time from $t_x := t_{N-1}$ to 
the corresponding $t_*$. Thereafter, we are ready to add another full
hexagon which ensures the limiting behaviour.
With $t_x=t_0\cdot1/(1-\tilde S(M+1))=t_0/((M+1)(1-S))$ and
$t_*=t_0\cdot(1-I)/(1-S)$, we obtain
${t_*}/{t_x} = (M+1)\cdot (1-I)$
for the ``half'' hexagon and a total time of
$$
t_{N-1}\cdot (M+1)\cdot (1-I)\cdot \frac{1-I}{1-S}=
t_{N-1}\cdot (M+1)\cdot \frac{(1-I)^2}{1-S}
$$
to reach the end of the full hexagon.
During the hexagon phase, 
we allow only a single extension (putting $\delta=1$, whenever
$\delta\neq 0$ is required) and thus produce a single well-defined 
discrepancy sequence. We then still have 
$$
\left(q^M\right)^{t_{N-1}\cdot \left(1-\frac{1}{N}\right)}\cdot
1^{t_{N-1}\left((M+1)\frac{(1-I)^2}{1-S}-1\right)}
= \left[\left(q^M\right)^{t_{N-1}\cdot
    (M+1)\frac{(1-I)^2}{1-S}}\right]
^{\left(\frac{(1-1/N)(1-S)}{(M+1)(1-I)^2}\right)}
$$
prefixes of length ${t_{N-1}\cdot (M+1)\frac{(1-I)^2}{1-S}}$ in 
${\cal  A}(I,S)$, which leads to a  Hausdorff
dimension at least $\frac{(1-1/N)(1-S)}{(M+1)(1-I)^2}$.

By \cite{NW,WN}, almost all sequences in
$\left(\ff_q^{K'}\right)^\infty$  lead to $I=S=\frac{M}{M+1}$ or $\tilde
I=\tilde S=0$ and thus can be used to fill between haxagons without
leaving the bounds $I$ and $S$. Hence it is possible to reach some
$t_N$ at the end of fill $F_N$ with at least 
$\left(q^M\right)^{t_N\cdot (1-\frac{1}{N-1})}$ different prefixes. 
The Hausdorff dimension of ${\cal A}(I,S)$ thus is lowerbounded by the 
number of prefixes at the end of the hexagons, with $n\to \infty$ thus
$$D_H\geq \frac{1-S}{(M+1)(1-I)^2}.$$

Finally, with $M>K'$, only $K'$ sequences may be active, 
the other $M-K'$ depending
$\ff_q(x)$--linearly on them. Letting the first $K'$
sequences fix $I$ and $S$, gives as before 
$\frac{1-S}{(M+1)(1-I)^2}\leq D_H\leq 1$ 
in $\left(\ff_q^{K'}\right)^\infty$ and thus\linebreak
$\frac{K'}{M}\cdot\frac{1-S}{(M+1)(1-I)^2}\leq D_H\leq \frac{K'}{M}$
in $\fqw$. The remaining sequences are $\ff_q(x)$--dependent,
hence increase the number of feasible sequences only by a 
factor of ${M \choose K'}\cdot
|\ff_q(x)|^{(M-K')\cdot K'}=\aleph_0$, too few to change~$D_H$.
\end{proof}

\section*{Conclusion}

We have determined all possible values for the asymptotic behaviour of the
normalized linear complexity of multisequences. We have also given an
algorithm to actually produce a sequence of any multiness $M$ 
with prescribed infimum $I$ and
supremum $S$ of its normalized linear complexity. This gives a
positive  answer to the
question posed by Dai, Imamura and Yang, whether the well--known
equality $\liminf L_a(n)/n + \limsup L_a(n)/n = 1$ in the case of
one sequence has a generalization.

We finished with the cardinality, Hausdorff dimension, and measure of
the set ${\cal A}(I,S)$ of sequences attaining  the prescribed bounds,
obtaining that all sets ${\cal A}(I,S)$ have $2^{\aleph_0}$ elements,
and, at least for $S\neq 1$,  positive Hausdorff dimension.

\end{document}